# Economic Evaluation of V2G-Enabled Fast Charging Stations Under Endogenous EV Adoption Dynamics

Mingjian Tuo[1,3][0000-0002-3942-3188] (✉), Jie Zhou[2], Yao Yan[2], Cunzhi Zhao[4], Long Wang[2], Mulan Zhang[3]

[1] Hubei Key Laboratory of Energy Storage and Power Battery (Hubei University of Automotive Technology), Shiyan, Hubei Province 442000, China
[2] Hubei University of Automotive Technology, Shiyan, Hubei Province 442000, China
[3] Key Laboratory of Automotive Power Train and Electronics (Hubei University of Automotive Technology), Shiyan, Hubei Province 442000, China
[4] Department of Engineering and Computer Science, McNeese State University, Lake Charles, Louisiana, LA, 70605, USA
mtuo@huat.edu.cn

**Abstract.** Building fast charging stations (FCSs) is crucial for transportation electrification, but there exists an indirect network effect: while the increasing number of electric vehicles (EVs) decides the FCS capacity expansion, the spatial locations of these facilities strongly influence drivers' willingness to adopt EVs. Ignoring this interaction can lead to bad capital investments and exacerbate power grid vulnerabilities during tidal traffic peaks. Therefore, we explicitly model the EV adoption dynamics as decision-dependent uncertainties (DDUs) in a new multi-period collaborative planning framework. This framework evaluates the economic viability of V2G-enabled FCSs across both transportation and distribution networks. To simplify the complex calculation, we introduce an aggregated fleet virtual battery model to catch macroscopic vehicle-to-grid (V2G) flexibility. This successfully circumvents the dimension curse inherent in tracking microscopic state-of-charge. To further guarantee calculation speed, the nonlinear infrastructure exposure is transformed into a mixed-integer program by using Special Ordered Set type 2 (SOS2) variables and Second-Order Cone Programming (SOCP) relaxations for grid limits. Finally, numerical studies on a coupled Sioux Falls and IEEE 33-bus testbed prove that our framework achieves superior expected social welfare. Also, macroscopic V2G aggregation is highlighted for its capability to mitigate distribution grid congestion penalties.

## 1 Introduction

Electric vehicles (EVs) are very important for the decarbonization of our modern energy systems. With more and more EVs on the road, fast charging stations (FCSs) have naturally become the most important physical interfaces that connect the power grid and transportation networks. Recently, many researchers have studied this connection from different angles, such as collaborative optimization [1], cross-system interactions [2], and robust scheduling [3]. At the same time, some other works try to improve charging guidance by nudging users [4] or quantifying their willingness [5].

However, we found a critical research gap: most current studies just assume the EV demand is exogenous [6-7]. This kind of simplification completely cuts off the endogenous feedback loop. Actually, building more charging infrastructure will directly encourage more people to buy EVs [8]. Because uncoordinated charging can cause serious grid congestion and voltage violations [9-10], if we ignore this endogenous dynamic, we will probably make wrong capital investments. It will also make the power grid much more vulnerable, especially during the tidal traffic peaks.

To deal with this congestion problem, integrating vehicle-to-grid (V2G) flexibility is a very practical method. Today, responsive V2G hardware is becoming a reality through AC/DC hybrid microgrids [11]. Also, there are many operational improvements like robust voltage algorithms [12] and predictive control. But here comes a new challenge: if we want to explicitly track the microscopic state-of-charge (SOC) for thousands of moving vehicles, the computational burden will be too huge and totally intractable for our macroscopic planning models.

Therefore, to fill these research gaps, this paper proposes a unified collaborative planning framework for V2G-enabled FCSs. This framework is explicitly driven by the endogenous EV adoption dynamics. Our main contributions are as follows:

(1) Endogenous modeling of adoption dynamics: To make sure the model can be effectively solved by Mixed-Integer Linear Programming (MILP), we implement a data-driven piecewise linear mapping. This convexification helps us quantify the decision-dependent adoption trajectories, so we can capture the real economic benefits of infrastructure stimuli.

(2) Macroscopic V2G aggregation: We build an aggregated fleet virtual battery model based on a macroscopic causal inequality. By doing this, we can extract the large-scale V2G arbitrage capabilities. More importantly, we completely avoid the combinatorial explosion caused by the microscale SOC tracking.

# 2 System Modeling

## 2.1 Endogenous EV Adoption Dynamics (DDU)

EV market penetration conforms to a logistic diffusion process, where the adoption saturation limit is endogenously governed by regional infrastructure deployment. The expected EV adoption rate $s_{od,t}$ for an origin-destination (OD) pair $od$ at macro-planning period $t$ is formulated as:

$$s_{od,t} = s_{od,t-1} + \alpha^{\text{base}}\left(S^{\text{sat}} - s_{od,t-1}\right) + \Delta s_{od,t}^{DDU}(X_t) \tag{1}$$

Directly embedding the nonlinear saturation increment $\Delta s_{od,t}^{DDU}(X_t)$ compromises the tractability of the Mixed-Integer Linear Programming (MILP) master problem. Therefore, it is reformulated via a piecewise linear mapping using Special Ordered Set type 2 (SOS2) variables:

$$\Delta s_{od,t}^{DDU} = \sum_{m \in \mathcal{M}} \alpha_{od,t,m}^{SOS2} \cdot \Psi_m^{DDU} \tag{2}$$

$$\sum_{m \in \mathcal{M}} \alpha_{od,t,m}^{SOS2} = 1, \quad \alpha_{od,t,m}^{SOS2} \geq 0, \quad \forall od, t \tag{3}$$

where $\alpha_{od,t,m}^{SOS2}$ represent continuous interpolation weights subject to SOS2 adjacency constraints. This geometric projection translates the decision-dependent behavioral feedback into a strictly polyhedral feasible region, ensuring exact compatibility with commercial branch-and-cut solvers.

### 2.2 Spatiotemporal Traffic Flow and Virtual Queuing

Heterogeneous urban mobility is captured by projecting the macroscopic EV adoption trajectory onto a 24-hour operational horizon via destination-based tidal probability distributions $\rho_{od,h}^{\text{zone}}$:

$$D_{od,t,h} = s_{od,t} \cdot D_{od}^{\text{base}} \cdot \rho_{od,h}^{\text{zone}} \tag{4}$$

Standard dynamic traffic assignments mapping to candidate stations entail combinatorial explosion due to microscopic SOC tracking. As a computationally tractable alternative, we implement a static spatial topology mask bounded by SOC continuity limits (e.g., maximal physical driving range). By introducing a virtual queuing state variable $Q_{od,t,h}^{shift}$ to buffer temporal congestion, load shedding triggers are bypassed, and the traffic assignment reduces to a linear mapping:

$$\sum_{i \in \mathcal{K}_{od}} \lambda_{od,i,t,h} + \lambda_{od,t,h}^{\text{un}} + Q_{od,t,h}^{\text{shift}} = D_{od,t,h} + Q_{od,t,h-1}^{\text{shift}} \tag{5}$$

A terminal queue clearance constraint prevents inter-day unserved demand deferral:

$$Q_{od,t,24}^{shift} = 0 \tag{6}$$

### 2.3 Active Distribution Network and Macroscopic V2G

Evaluating the operational feasibility of the power grid requires a computationally efficient formulation. We formulate the lower-layer Optimal Power Flow (OPF) as a penalty-minimization primal problem:

$$\min \sum_{h \in \mathcal{H}} \Bigg( c_{\text{imp}} P_{root,h}^{\text{import}} + c_{\text{loss}} \sum_{(n,m) \in \mathcal{L}} R_{nm}\, l_{nm,h} + c_{\text{deg}} \sum_{m \in \mathcal{N}} P_{m,h}^{V2G}$$

$$+c_{\text{shed}}\sum_{m\in\mathcal{N}}P_{m,h}^{\text{shed}}+c_{\text{pen}}\Phi_h^{\text{grid}}\Bigg) \quad (7)$$

This objective explicitly incorporates the active power import cost $c_{imp}$, system losses, battery degradation $c_{\text{deg}}$, and load shedding $c_{shed}$. The term $\Phi_h^{\text{grid}}$ encapsulates all auxiliary slack variables (e.g., voltage deviations and thermal overloads) to prevent algorithmic deadlocks.

Governed by this objective, we adopt the DistFlow branch model to formulate the distribution network operations:

$$\sum_{k\in\delta(m)}P_{mk,h}-\sum_{n\in\pi(m)}P_{nm,h}+R_{nm}l_{nm,h}=P_{m,h}^{DG}$$

$$+P_{m,h}^{V2G}+P_{m,h}^{\text{shed}}-P_{m,h}^{\text{load}}-P_{m,h}^{EV} \quad (8)$$

$$\sum_{k\in\delta(m)}Q_{mk,h}-\sum_{n\in\pi(m)}Q_{nm,h}+X_{nm}l_{nm,h}=Q_{m,h}^{DG}-Q_{m,h}^{\text{load}}+Q_{m,h}^{EV} \quad (9)$$

$$v_{m,h}=v_{n,h}-2\big(R_{nm}P_{nm,h}+X_{nm}Q_{nm,h}\big)+(R_{nm}^2+X_{nm}^2)l_{nm,h}+v_{m,h}^{\text{slack}} \quad (10)$$

where $P_{nm,h}$ and $Q_{nm,h}$ are the active and reactive branch flows, $v_{m,h}$ and $l_{nm,h}$ are the squared nodal voltage and branch current magnitudes. To eliminate the AC-OPF non-convexities, the Second-Order Cone Programming (SOCP) relaxation is applied to the branch power boundaries:

$$P_{nm,h}^2+Q_{nm,h}^2\le l_{nm,h}v_{n,h} \quad (11)$$

Crucially, to abstract the V2G flexibility of the fleet without tracking individual plug-in states, a macroscopic causal inequality is established. The continuous variables representing V2G discharging support $P_{m,h}^{V2G}$ and charging extraction $P_{m,h}^{EV}$ are bounded by a rolling multi-period integration constraint:

$$\sum_{\tau=1}^{h}\frac{1}{\eta_d}P_{m,\tau}^{V2G}\cdot\Delta h\le\sum_{\tau=1}^{h}\eta_c\,P_{m,\tau}^{EV}\cdot\Delta h+E_{m,0}^{\text{buffer}}+SOE_{m,h}^{\text{panic}} \quad (12)$$

This formulation guarantees that the cumulative V2G discharging energy never exceeds the sum of the cumulative charging energy and the initial fleet energy cushion $E_{m,0}^{\text{buffer}}$, ensuring aggregate thermodynamic consistency.

## 3 Problem Formulation

The co-planning framework is formulated to optimally balance capital expenditure (CAPEX) and multi-sector operational reliability. The master objective function minimizes the equivalent annualized investment cost alongside the expected operational penalty costs:

$$\min_{z,x} F^{UP} = \frac{F_{\text{inv}}(z,x)}{\eta^{LCA}} - F^{\text{revenue}}(x) + F_{\text{pen}}^{\text{traf}}(x) + F_{\text{pen}}^{\text{grid}}(x) \tag{13}$$

Here, $F_{\text{inv}}$ encapsulates the fixed construction and modular procurement costs. $F^{\text{revenue}}$ denotes the social service dividends generated by fulfilling EV charging demands, shifting the paradigm from passive cost-minimization to active welfare-maximization. $F_{pen}^{traf}$ and $F_{pen}^{grid}$ represent the exact economic penalties for traffic shedding/queuing and distribution grid violations, respectively, evaluated over representative spatiotemporal profiles.

# 4 Case Study

## 4.1 System Setup and Parameters

The proposed physical framework is evaluated on a tightly coupled testbed comprising the 24-node Sioux Falls transportation network and the IEEE 33-bus active distribution system. The macro-planning horizon spans 6 years, divided into 3 investment stages. To accurately reflect real-world economic trade-offs, the Value of Lost Load (VOLL) for the distribution grid is set to 25,000 \$/MWh to penalize operational violations, while the unserved EV traffic penalty is set to 500 \$/MWh. The single-module fast-charging capacity is 120 kW.

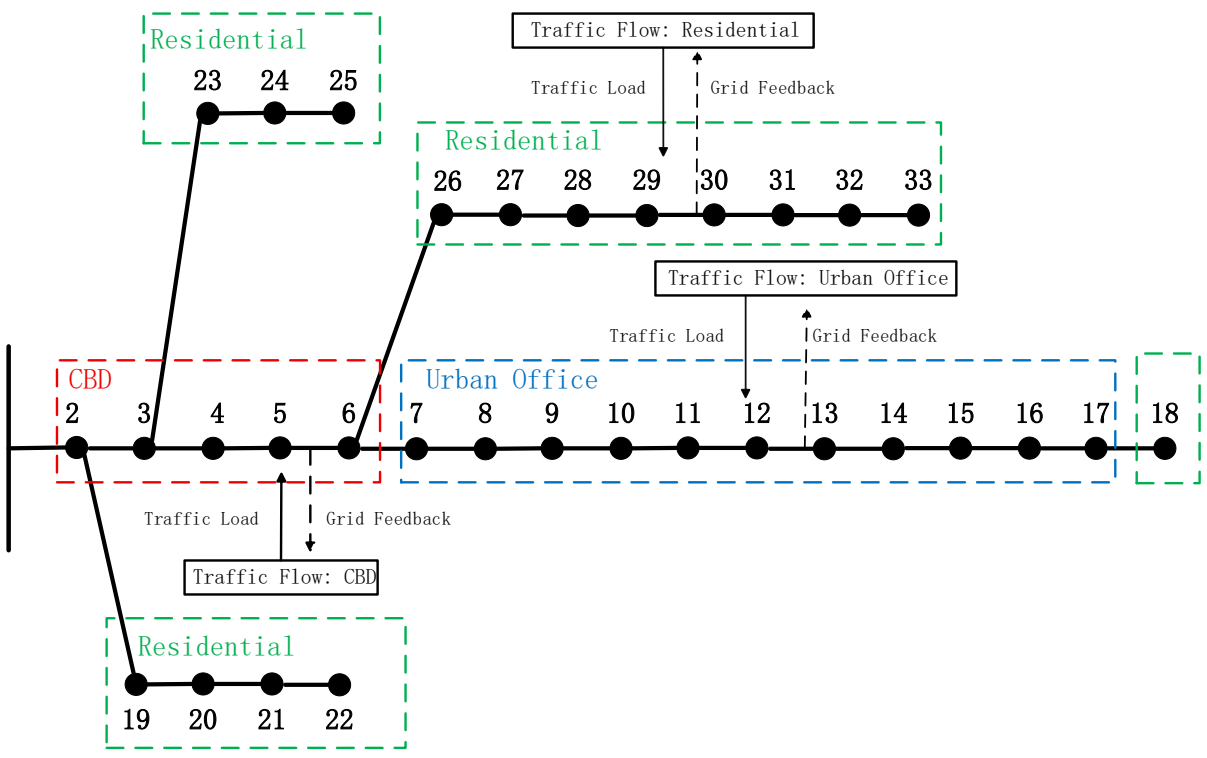

**Fig. 1.** Spatial coupling topology of the IEEE 33-bus system and the transportation network.

As illustrated in Fig. 1, to capture the heterogeneous spatial coupling, dynamic traffic nodes are mapped to the distribution grid based on their functional centrality. High-traffic commercial business districts (CBDs) are anchored to the robust main trunk buses (Buses 2-6) to withstand tidal charging shocks, whereas urban office and residential nodes are allocated to secondary (Buses 7-17) and leaf feeders (Buses 18-33), respectively.

### 4.2 Economic Value of Endogenous DDU Modeling

The economic and physical performance of the proposed formulation is benchmarked against three alternative planning paradigms under identical initial conditions: Strategy I (Bidirectional V2G + Endogenous DDU), Strategy II (Endogenous DDU only), Strategy III (Bidirectional V2G only), and Strategy IV (Traditional Open-Loop).

**Table 1.** Comparison of planning strategies regarding economic and physical performance.

| Strategy | Expected EV Adoption (%) | Expected Service Coverage (%) | Grid Violations (p.u./MWh) | Total CAPEX (M$) | V2G Energy Ratio (%) | Apparent Welfare (In-Sample M$) |
|---|---|---|---|---|---|---|
| Strategy_I | 25.36 | 100 | 217.46 | 57.37 | 91.18 | 227.93 |
| Strategy_II | 26.69 | 100 | 15.75 | 56.92 | 0 | 224.27 |
| Strategy_III | 19.88 | 99.99 | 215.14 | 49.47 | 91.23 | 268.86 |
| Strategy_IV | 15.37 | 100 | 15.75 | 45.37 | 0 | 247.45 |

Table 1 reveals some profound and somewhat counter-intuitive insights. Actually, the traditional open-loop paradigm (Strategy IV) severely underestimates the EV adoption rate (only reaching 15.37%) because it simply treats charging demand as a static exogenous parameter. As a result, it leads to a highly conservative capacity deployment with the lowest CAPEX of 45.37 M$.

On the contrary, applying the DDU mapping (Strategies I and II) inherently clusters capacity expansion around core commercial hubs. This directly accelerates EV adoption to 25.36% and 26.69%. We have to admit this imposes a heavier CAPEX (57.37 M$ under Strategy I). However, the endogenously stimulated demand effectively offsets these initial outlays, ultimately unlocking huge long-term social service benefits.

What's more, there is a distinct physical trade-off in the distribution network's utilization. Implementing V2G (Strategies I and III) drives grid congestion indices to about 215–217 p.u./MWh, while non-V2G setups entirely avoid this. Fundamentally, V2G energy arbitrage takes up a massive margin of grid capacity. Strategy I mathematically balances this well, successfully maintaining a 25.36% EV penetration while capturing 91.18% of the V2G potential. But Strategy III is totally different. It depletes grid capacity purely for energy arbitrage, leaving EV adoption stagnated at 19.88%. Therefore, we can get a definitive conclusion: deploying V2G without considering endogenous traffic feedback leads to fundamentally inefficient infrastructure utilization.

### 4.3 Value Breakdown of V2G and Closed-Loop Planning

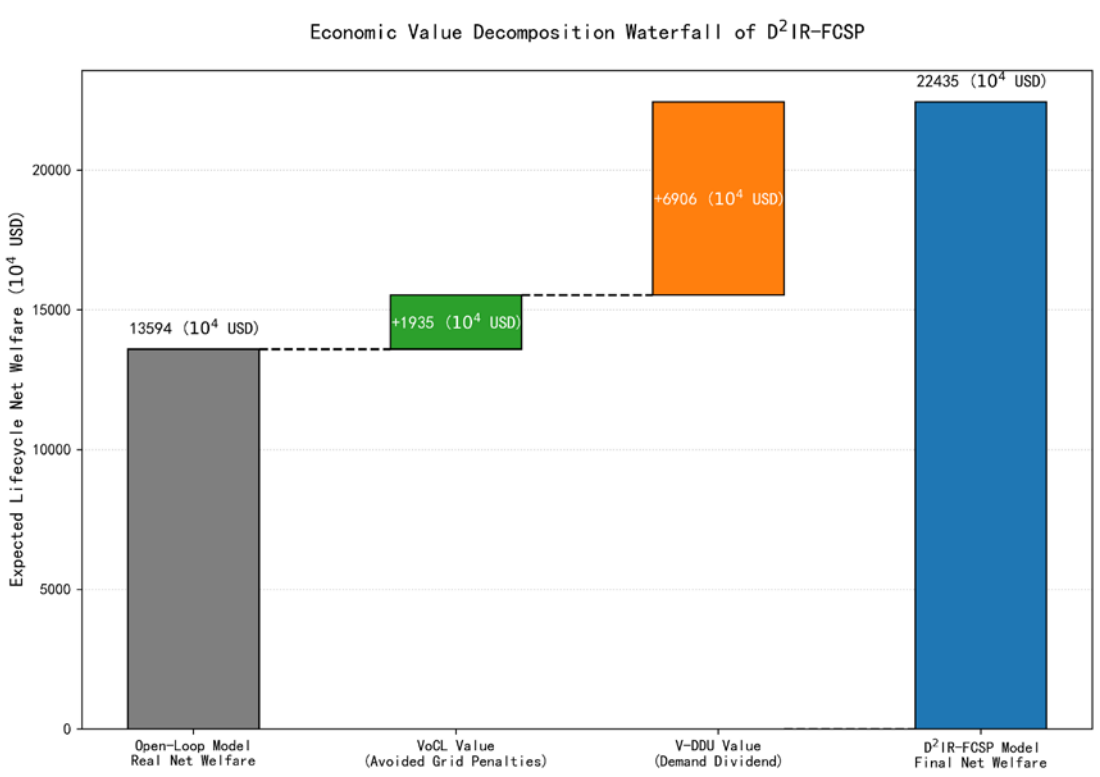

**Fig. 2.** Breakdown of economic drivers and synergistic benefits via a waterfall chart.

To clearly evaluate the economic drivers behind our optimal strategy, Fig. 2 uses a waterfall chart to illustrate the decomposition of synergistic dividends. Obviously, the welfare improvement from the Strategy IV (our baseline) to the Strategy I (the global optimum) is mainly because of two important components:

Decision-related uncertainty value (V-DDU): Traditional exogenous models always have some inherent forecasting deviations. By eliminating these deviations, our proposed framework successfully catches the extra revenue generated by the infrastructure-stimulated EV fleet. As a result, this endogenous interaction finally leads to an absolute market penetration increase of about 10%.

Closed-loop V2G value (VoCL): Instead of passively accepting the tidal traffic congestion, our framework takes full advantage of the 91.18% V2G energy capacity, treating it as a huge aggregated virtual battery. After that, spatiotemporal energy arbitrage is utilized to offset the heavy 57.37 M$ capital expenditure (CAPEX). Therefore, the active distribution network is not just a strict physical constraint anymore. It now functions proactively to help us maximize the overall social welfare.

## 5 Conclusion

To sum up, this paper proposes a collaborative planning framework for V2G-enabled fast charging stations, which is explicitly driven by endogenous EV adoption dynamics. To bypass the terrible combinatorial explosion caused by microscopic vehicle tracking, we formulate a macroscopic virtual battery causal inequality. This method successfully guarantees computational tractability while rigorously preserving exact V2G arbitrage boundary conditions. Through our numerical simulations, the superiority and feasibility of the proposed framework are well validated. The results clearly demonstrate that coupling decision-dependent uncertainty (DDU) mapping with macroscopic V2G dispatch not only accelerates regional EV adoption but also completely neutralizes uncoordinated grid congestion. What's more, the exact convexification established here provides a strict mathematical baseline. Building upon this solid foundation, our future research will explore large-scale heuristic decomposition and out-of-sample robustness verification in coupled power-traffic systems.